# Self-consistent theory of reversible ligand binding to a spherical cell

Shivam Ghosh(1,2), Manoj Gopalakrishnan* (1) and Kimberly Forsten-Williams(3)

1 Harish-Chandra Research Institute, Jhunsi, Allahabad 211019, India.
2 Department of Physics, St Stephen's College, University Enclave, Delhi 110 007, India.
3 Department of Chemical Engineering, Virginia Polytechnic Institute and State University, Blacksburg, VA 24061, USA.

**Abstract.** In this article, we study the kinetics of reversible ligand binding to receptors on a spherical cell surface using a self-consistent stochastic theory. Binding, dissociation, diffusion and rebinding of ligands are incorporated into the theory in a systematic manner. We derive explicitly the time evolution of the ligand-bound receptor fraction p(t) in various regimes . Contrary to the commonly accepted view, we find that the well-known Berg-Purcell scaling for the association rate is modified as a function of time. Specifically, the effective on-rate changes non-monotonically as a function of time and equals the intrinsic rate at very early as well as late times, while being approximately equal to the Berg-Purcell value at intermediate times. The effective dissociation rate, as it appears in the binding curve or measured in a dissociation experiment, is strongly modified by rebinding events and assumes the Berg-Purcell value except at very late times, where the decay is algebraic and not exponential. In equilibrium, the ligand concentration everywhere in the solution is the same and equals its spatial mean, thus ensuring that there is no depletion in the vicinity of the cell. Implications of our results for binding experiments and numerical simulations of ligand-receptor systems are also discussed.

## 1. Introduction

A large number of cellular processes are triggered by ligand binding to cell surface receptors. In contrast to reactions taking place within a test tube, ligand-receptor cell binding involves reactants that are not `well-mixed', primarily because the receptors are confined to the cell surface. In addition, the ligand concentration close to the surface is subject to fluctuations arising from association and dissociation processes. The kinetics of ligand-receptor binding could, therefore, deviate substantially from the traditional ordinary differential equation (ODE) based descriptions, which are appropriate for well-mixed reactants whose densities remain constant in space [1].

A related issue is the history-dependence of the binding process arising from the re-attachment of dissociated ligands to the cell surface at later times. For this reason, any effective equation describing the binding kinetics has to be necessarily non-local in time, an interesting feature which does not seem to have been widely appreciated in the literature. The traditional approach, starting with the classic paper on bacterial chemo-sensing by Berg and Purcell [2] and later extended by other authors [3,4,5] has been to find the effective rate constants by solving for the steady state ligand flux to the cell surface. Berg and Purcell showed that, for a spherical cell of radius $a$, with $N$ receptors

---

* Corresponding author. E-mail: manoj@mri.ernet.in

Self-consistent theory of ligand binding on its surface, each binding the ligand with an intrinsic on-rate $k_+$, the net ligand flux (without dissociation) at the surface is $J \equiv k_f \rho_0$, where $\rho_0$ is the concentration at infinity and

$$k_f = 4\pi a D N k_+ /(N k_+ + 4\pi a D) \qquad (1)$$

is the 'effective' on-rate for the process and $D$ is the diffusion coefficient for the ligand. DeLisi and Wiegel [3], and later, Shoup and Szabo [4] showed that the off-rate also undergoes a similar scaling because of rebinding of ligands to the surface, following dissociation at early times. Shoup and Szabo called the factor

$$\gamma = \frac{N k_+}{4\pi a D + N k_+} \qquad (2)$$

the absorption probability of a ligand upon contact (and, consequently $1-\gamma$ is the probability of non-absorption). It then follows that the effective rate constants may be expressed as $k_f = 4\pi a D \gamma$ and $k_b = k_-(1-\gamma)$, which gives a nice heuristic interpretation of these results.

The scaling relations for the binding and dissociation rates mentioned above have been found to agree well with experimental results [6,7], are widely used in biochemical modeling, and have also been extended to cell surface reactions. In addition, Berg and Purcell's well-known result for the relative error in concentration measurement by a chemotactic cell (like the bacterium E. Coli) follows directly from Eq.1. It is, therefore, of interest to re-examine the validity of the quasi-steady state assumptions (where the ligand concentration is assumed to reach a stationary profile much faster than the receptors) under which these relations were derived and look for possible modifications in a more systematic theory. In addition, we note that in the previous theoretical treatments, association and dissociation processes are treated separately, and, in particular, effects of ligand rebinding on the association kinetics are neglected altogether [8,9]. An attempt towards a unified treatment of binding and dissociation was made by Goldstein and Dembo [10]. However, this technique (method of weighted residuals) is essentially perturbative in nature and does not always predict the correct kinetic behavior. For example, in the pure dissociation problem, the asymptotic decay is algebraic, but MWR predicts exponential decay[10].

In this paper, we generalize a recently introduced self-consistent stochastic formalism [11,12] originally developed to study ligand rebinding to uniformly distributed receptors on a planar surface[12], and later extended to include receptor clusters[11], to study association and rebinding of ligands to receptors on a spherical cell in solution. Being based on the statistics of individual ligand trajectories, this formalism eliminates the need to impose ad hoc boundary conditions for the variation of ligand concentration close to the cell surface. We derive the ligand-bound receptor fraction explicitly as a function of time, and thus derive the effective rate constants. We observe deviations from Berg-Purcell scaling close to equilibrium, and show that the association rate equals the intrinsic rate (and the ligand concentration at the surface equals its spatial mean value on solution) at equilibrium. We discuss experimental implications of our findings, and also test the impact of our predictions in a simple computational model based on the epidermal growth factor and receptor (EGF-EGFR) system.

## 2. Self-consistent theory of binding

*2.1 General formalism*

Let us consider a spherical cell of radius $a$ in three dimensions, with $N = 4\pi a^2 R_0$ receptors distributed randomly on its surface, $R_0$ being the surface density. The cell is exposed to a ligand in the surrounding medium, whose concentration at $t=0$ and at infinite distance from the cell is $\rho_0$. Let us denote by $p(t)$ the fraction of ligand-bound receptors at time $t$. The basic kinetic equation for $p(t)$ is

$$\frac{dp(t)}{dt} = -k_- p(t) + k_+ \rho(a,t)[1-p(t)] \tag{3}$$

where $\rho(r,t)$ is the radially symmetric ligand concentration at distance from the center of the cell, at time $t$. $k_-$ and $k_+$ are the dissociation and association rates of the ligand to the receptor. For the rest of this section, we will assume that internalization of ligand-receptor complexes by the cell can be neglected in comparison with dissociation.

It is convenient to write $\rho(a,t)$ in the form $\rho(a,t) = \rho_b(a,t) + \rho_r(a,t)$. $\rho_b(a,t)$ is the density contribution 'from the bulk', i.e., from the ligands which have not been bound to a receptor until time $t$ (although they may have been 'reflected' earlier by bound-receptors or the non-receptor part of the surface). The ligands that make up $\rho_r(a,t)$ are those which were receptor-bound at an earlier time $\tau < t$ and released during the interval $[\tau : \tau + d\tau]$. Let us now define a Green's function for ligand diffusion around the semi-absorbing sphere centered at $r=0$: $G_N(a,t;r,0) \equiv G_N(r,t)$ is the probability density (dimension 1/length$^3$) that a ligand that starts at a radial distance $r$ from the origin at $t=0$ is located close to the surface at time $t$, with possibly multiple visits to the semi-absorbing surface in between (but no absorption by a receptor). Using this Green's function, the above two densities may now be expressed as

$$\rho_b(a,t) \equiv \rho_b(t) = \rho_0 \int_a^\infty G_N(r,t) 4\pi r^2 dr \tag{4a}$$

$$\rho_r(a,t) \equiv \rho_r(t) = k_- N \int_0^t d\tau p(\tau) G_N(a,t-\tau) \tag{4b}$$

The first relation is obvious. In order to understand the second relation, let us assume the ligand under consideration was bound to a receptor last at time $\tau$ and was released during the interval $[\tau : \tau + d\tau]$, which takes place with probability $k_- d\tau$. $Np(\tau)$ is the total number of bound receptors at time $\tau$, and the probability that the ligand returns to the surface in the interval $[\tau : t]$ without being absorbed by a receptor in between is $G_N(a,t-\tau)$. Note that we have assumed complete spherical symmetry of ligand diffusion, which is justified for a uniform distribution of receptors as is the case here.

Self-consistent theory of ligand binding

The separation of the surface ligand density into bulk and rebinding-originated parts is the first new feature of our formalism. The second feature is the expression of these densities in terms of Green's function for individual ligand diffusion trajectories (in the presence of a semi-absorbing surface). The Green's function is computed in the next sub-section.

*2.2 Calculation of $G_N(r,t)$*

The formalism in this section is very similar to what was presented in our earlier paper [11]. Let us first define the first passage probability density $q(r,t)$, which is the probability (per unit volume) that a ligand at radial distance $r$ at t=0 arrives close to the surface for the first time at time t, which is related to $G_N(r,t)$ as

$$G_N(r,t) = q(r,t) + \eta \int_0^t \frac{d\tau}{\delta} \lambda q(r,\tau) G_N(a,t-\tau) 4\pi a^2. \tag{5}$$

Here, $\lambda$ is a microscopic length scale and $\delta$ is the time interval for which a ligand will reside inside a volume element $\lambda^3$, before moving away by diffusion. The first term gives the probability that the ligand arrives at the surface for the first time at t, the second term gives the sum of all other events. The factor $\eta = 1 - k_+ R_0 \delta / \lambda$ gives the probability of non-binding upon contact with the surface (In general, close to equilibrium, we need to correct for the number of occupied receptors, which can be done approximately by replacing $R_0$ by $R_0(1-p_s)$ in $\eta$. See also the note following Eq.9b).

In order to calculate $q(r,t)$, we next consider the case $R_0 = 0$ so that the surface is completely reflecting as far as ligand diffusion is concerned. In this case, $\eta = 1$ and hence Eq.5 reduces to

$$G_0(r,t) = q(r,t) + \frac{\lambda}{\delta} \int_0^t d\tau q(r,\tau) G_0(a,t-\tau) 4\pi a^2 \tag{6}$$

We will now re-express Eq.5 and 6 in terms of Laplace-transforms $\tilde{q}(r,s)$ and $\tilde{G}_N(r,s)$, and eliminate $\tilde{q}(r,s)$ between the equations, which produces the elegant relation

$$\tilde{G}_N(r,s) = \frac{\tilde{G}_0(r,s)}{1 + k_+ N \tilde{G}_0(a,s)}. \tag{7}$$

The calculation of $\tilde{G}_0(r,s)$ can be done using standard techniques and is reproduced for easy reference in Appendix A. The result is

Self-consistent theory of ligand binding

$$\widetilde{G}_0(r,s) = \frac{1}{4\pi a r \sqrt{Ds}} \left[1 + \sqrt{D/a^2 s}\right]^{-1} e^{-(r-a)\sqrt{s/D}} \tag{8}$$

After substituting Eq.8 into Eq.7 and using the basic equations Eq.4a and 4b, we arrive at the following expressions for the Laplace transforms of the densities: For the bulk density, we find

$$\widetilde{\rho}_b(s) = \frac{\rho_0}{s}\left[\frac{1+a\sqrt{s/D}}{(1-\gamma)^{-1} + a\sqrt{s/D}}\right], \tag{9a}$$

where the factor $\gamma$ is defined in Eq.2. Inversion [13] gives

$$\rho_b(t) = \rho_0\left[1-\gamma + \gamma e^{\frac{Dt}{a^2(1-\gamma)^2}} \text{erfc}\left[\frac{\sqrt{Dt}}{a(1-\gamma)}\right]\right] \approx \rho_0(1-\gamma) + \frac{\gamma(1-\gamma)a}{\sqrt{\pi Dt}} \tag{9b}$$

where the last expression illustrates the approach to the steady state value at large times, $t \gg a^2(1-\gamma)/D$. The density contribution from receptor-released ligands (after using the theorem for convolutions in Eq.4b) is given by

$$\widetilde{\rho}_r(s) = \frac{Nk_-\widetilde{p}(s)}{4\pi aD(1+a\sqrt{s/D}) + Nk_+} \tag{9c}$$

For kinetics *close to the steady state*, the factor $N$ in Eq.2 and the denominators of Eq.7, Eq.9a-c should be corrected for the number of ligand-bound receptors, which is done by replacing $N$ by $N(1-p_s)$ in these places (See Eq.18 and Eq.19 later).

**3. Results**

The kinetics in various time regimes may now be worked out by inverting Eq.9a and 9c and substituting in Eq.3. We now study three important cases of interest.

*3.1 Early time regime, $t \ll a^2/D$:*

This is equivalent to the regime $a\sqrt{s/D} \gg 1$, for which Eq.9a gives $\widetilde{\rho}_b(s) \approx \frac{\rho_0}{s}$ so that $\rho_b(t) \approx \rho_0$, which is to be expected. In this limit, $\rho_r(t)$ is negligible since release and rebinding events are significant only after a considerable fraction of receptors are bound to ligands. The association curve in this regime is, therefore,

$$p(t) \approx \frac{\rho_0}{\rho_0 + K_d}\left[1 - e^{-(k_+\rho_0 + k_-)t}\right] \qquad t \ll a^2/D \tag{10}$$

Self-consistent theory of ligand binding where $K_d = k_-/k_+$. This regime is of very little practical interest since the time-scale $a^2/D$ is of the order of $10^{-3}$ seconds only (for $a \approx 5 \times 10^{-4}$ cm and $D \sim 10^{-5}$ cm$^2$ s$^{-1}$).

*3.2. Intermediate regime ($a^2/D \ll t, p(t) \ll 1$):*

In this case, we consider the regime beyond the (microscopic) time scale $a^2/D$, but the receptor occupancy by ligands is still small, i.e., $p(t) \ll 1$. In this case, using the condition $a\sqrt{s/D} \gg 1$, we find from Eq.9a that $\rho_b(t) \approx (1-\gamma)\rho_0$, i.e., the bulk-contributed density is now only a fraction of the (constant) density at infinity. Eq.9b can also be inverted trivially in this regime to give

$$\rho_r(t) \approx \frac{Nk_- p(t)}{4\pi aD + Nk_+} \quad \text{for } t \gg a^2/D. \tag{11}$$

Let us now substitute for $\rho_b(t)$ and $\rho_r(t)$ in Eq.3. The resulting equation is

$$dp(t)/dt \approx k_+(1-\gamma)\rho_0 - (1-\gamma)[k_+\rho_0 + k_-]p - \gamma k_- p^2 \tag{12}$$

For consistency, we neglect the quadratic term in p(t), and to linear order we have

$$p(t) \approx \frac{\rho_0}{\rho_0 + K_d}\left[1 - e^{-(1-\gamma)[k_+\rho_0 + k_-]t}\right] \quad a^2/D \ll t, p(t) \ll 1 \tag{13a}$$

which is characterized by a time scale

$$T_1 = \left[(1-\gamma)(k_+\rho_0 + k_-)\right]^{-1} \tag{13b}$$

We note from Eq.13 that in this regime, time is 'inflated' by a factor $(1-\gamma)^{-1}$, which may also be effectively absorbed into the reaction rates to define effective forward and backward rates

$$k_f = (1-\gamma)k_+ \text{ and } k_b = (1-\gamma)k_-. \tag{14}$$

Note that these effective rates are the same as predicted by the quasi-steady state arguments discussed in the introduction [2,4].

*3.3. Late time regime and steady state ($a^2/D \ll t, p(t) \sim p_s$):*

We now proceed to analyze the kinetics in the asymptotic regime, when the system is close to equilibrium. Let us first assume that at very late times, the quantities $p(t), \rho_b(t), \rho_r(t)$ assume their steady state values $p_s, \rho_b^s, \rho_r^s$ respectively. Let us therefore consider small deviations of the form

Self-consistent theory of ligand binding

$$p(t) = p_s + \delta p(t), \quad \rho_b(t) = \rho_b^s + \delta\rho_b(t), \quad \rho_r(t) = \rho_r^s + \delta\rho_r(t) \tag{15}$$

close to the steady state. In terms of Laplace-transformed variables, the equivalent relations are

$$\tilde{p}(s) = s^{-1}p_s + \delta p(s), \tilde{\rho}_b(s) = s^{-1}\rho_b^s + \delta\rho_b(s), \quad \tilde{\rho}_r(s) = s^{-1}\rho_r^s + \delta\rho_r(s) \tag{16}$$

We now substitute these expressions in Eq.9a and Eq.9c and the following relations follow easily. The steady state bound fraction is

$$p_s = \frac{\rho_s}{\rho_s + K_d}, \tag{17}$$

where the equilibrium surface concentration is $\rho_s = \rho_b^s + \rho_r^s$, with

$$\rho_b^s = (1-\gamma_s)\rho_0 \text{ and } \rho_r^s = \frac{Nk_- p_s}{4\pi aD + Nk_+(1-p_s)}. \tag{18}$$

Note that the scaling factor $1-\gamma_s$ appearing in Eq.18 is slightly different from the factor $1-\gamma$ in Eq.9a, and is defined as

$$\gamma_s = \frac{Nk_+(1-p_s)}{4\pi aD + Nk_+(1-p_s)}. \tag{19}$$

As mentioned earlier, this (approximate) modification reflects the reduction in the number of available sites for ligand association as binding progresses. Upon solving Eq.17 and Eq.18 together, it is easily seen that $\rho_r^s = \gamma_s \rho_0$ and hence $\tilde{\rho}_b(s) + \tilde{\rho}_r(s) = \rho_0$, i.e., *the ligand density close to the surface at equilibrium is the same as the bulk density*! (This is, in fact, true everywhere in the solution, see Appendix C). Note that this result is crucially dependent on the inclusion of the rebinding density $\rho_r$ and the correction factor $1-p_s$ in Eq.18 and Eq.19. From Eq.17 it also follows then that $p_s = \rho_0/(\rho_0 + K_d)$, i.e., the equilibrium bound fraction is insensitive to kinetic modifications of rate constants (ie., extrapolating Eq.10 and Eq.13 to $t \to \infty$ gives the same equilibrium value).

Having found the steady state in a precise manner, we now proceed to calculate the leading kinetic terms close to the steady state. After substituting Eq.15 into Eq.3, we find the following equation for $\delta p(t)$, up to linear terms in the deviations.

$$\frac{d\delta p(t)}{dt} \approx -(k_- + k_+\rho_0)\delta p(t) + k_+(1-p_s)[\delta\rho_b(t) + \delta\rho_r(t)]. \tag{20}$$

Eq.20 is Laplace-transformed as

Self-consistent theory of ligand binding

$$s\delta p(s) + p_s = -(k_- + k_+\rho_0)\delta p(s) + k_+(1-p_s)[\delta\rho_b(s) + \delta\rho_r(s)]. \tag{21}$$

Note that we have used a formal initial condition $\delta p(0) = -p_s$ however we recognize that our present analysis is valid only close to the steady state. In order to find $\delta\rho_b(s)$ and $\delta\rho_r(s)$, we first write $\tilde{\rho}_b(s) = \rho_b^s f_1(s)$ and $\tilde{\rho}_r(s) = \tilde{p}(s)f_2(s)$ where

$$f_1(s) = \frac{1 + a\sqrt{s/D}}{1 + a(1-\gamma_s)\sqrt{s/D}} \quad \text{and} \quad f_2(s) = \frac{Nk_-}{Nk_+(1-p_s) + 4\pi aD(1 + a\sqrt{s/D})}. \tag{22}$$

It then follows that

$$\delta\rho_b(s) = s^{-1}\rho_b^s[f_1(s) - 1] \quad \text{and} \quad \delta\rho_r(s) \approx f_2(s)\delta p(s) + s^{-1}p_s[f_2(s) - f_2(0)]. \tag{23}$$

In the next step, we expand the functions $f_1$ and $f_2$ around $s = 0$, in powers of the dimensionless variable $a\sqrt{s/D}$. It is then found that

$$f_1(s) \approx 1 + \gamma_s a\sqrt{s/D} - \gamma_s(1-\gamma_s)(a\sqrt{s/D})^2 + \ldots \tag{24a}$$
$$f_2(s) \approx f_2(0)\left[1 - (1-\gamma_s)a\sqrt{s/D} + (1-\gamma_s)^2(a\sqrt{s/D})^2 + \ldots\right] \tag{24b}$$

In the final step of the procedure, the expansions Eq.24a and Eq.24b are used in Eq.23, which is then substituted into Eq.21. After a few elementary simplifications, we arrive at the final result:

$$\delta p(s) \approx -p_s\left[s + k_+\rho_0 + k_-(1-\gamma_s)[1 + \gamma_s a\sqrt{s/D}]\right]^{-1} \tag{25}$$

We see that this regime is characterized by a time-scale

$$T_2 = (k_+\rho_0 + (1-\gamma_s)k_-)^{-1}. \tag{26}$$

Eq.25 may be explicitly inverted for two limiting cases. If $a\sqrt{s/D} \ll 1$, but $s \gg T_2^{-1}$ (equivalently, when $a^2/D \ll t \ll T_2$), we find $\delta p(t) \approx -p_s e^{-t/T_2}$, and hence

$$p(t) \approx p_s[1 - e^{-t/T_2}] \quad \text{when } a^2/D \ll t \ll T_2, \text{ and } p(t) \approx p_s \tag{27}$$

The time evolution in this regime is still exponential, with the time scale defined by Eq.26. Eq.26 and Eq.27 are the principal results of this paper. Note that, unlike the time scale appearing in Eq.13, in Eq.26 only the dissociation rate is scaled by the factor $1-\gamma_s$, whereas the association rate has now recovered to its original, intrinsic value.

Self-consistent theory of ligand binding

It would appear from our analysis that, unlike what is found using the standard Berg-Purcell-Shoup-Szabo rates (Eq.13), the equilibrium dissociation constant $K_d^{eff} \equiv k_r / k_f$, as measured from the binding curve, would be time-dependent, remaining at its intrinsic value $k_- / k_+$ over the early and intermediate regimes of evolution, but later settling down at a different value, smaller than the intrinsic value by a factor $1 - \gamma_s$. Note, however, that the steady state bound fraction still depends only on the intrinsic $K_d$ (Eq.17). This means that care should be taken when the effective rate constants, as extracted from kinetic measurements are used to predict quantities like the steady state receptor occupancy.

Although possibly of little experimental relevance, let us also take a look at the regime where $t \gg T_2$, which is arbitrarily close to equilibrium. The kinetics (which is now very slow, since we are very close to the steady state already) in this regime controlled by the $a\sqrt{s/D}$ term in Eq.25 since $s \ll T_2^{-1}$. Inversion[1] of Eq.25 and using the appropriate asymptotic expansion at large times [13] gives

$$\delta p(t) \cong -\frac{p_s T_2^2 a k_-(1-\gamma_s)}{2\sqrt{\pi D}} t^{-3/2} + \dots \quad \text{when} \quad t \gg T_2. \tag{28}$$

At very late times, therefore, the binding curve approaches its equilibrium value *algebraically*. The power-law approach to the final steady state is in agreement with a general result in the theory of bimolecular reactions, which predicts that, in $d$ dimensions, equilibrium is approached asymptotically as $t^{-d/2}$ [14,15,16]. The $t^{-3/2}$ mode of decay is also found, expectedly, in the pure dissociation problem (Eq.B6 in Appendix B).

Let us now restate our main results from a different perspective. The ligand concentration close to the surface varies non-monotonically with time: the bulk-originated contribution $\rho_b(t)$, after starting from $\rho_0$, decays to a fraction $(1-\gamma)\rho_0$ over a time scale $a^2/D$ (Eq.9a) and then slowly settles to its steady state value $(1-\gamma_s)\rho_0$. In contrast, $\rho_r(t)$, the rebinding contribution, is initially zero and, after a short transient $\sim a^2/D$, grows approximately proportional to $p(t)$ (Eq.11) and reaches equilibrium over a time scale $\sim T_2$ (Eq.26). The total surface concentration $\rho(t)$ therefore starts with the spatial mean value $\rho_0$ and quickly decays to $(1-\gamma)\rho_0$ due to absorption of ligands by the surface. But, as bound ligands are released, the density increases and asymptotically equals $\rho_0$. Of ourse, this is strictly true only in the absence of boundaries, so that the total number of ligands is infinite (while the total number of receptors remain finite) as has

---

[1] Inverse Laplace transform of $\tilde{g}(s) = (\sqrt{s} + a)^{-1}$ is $g(t) = (\pi t)^{-1/2} - \alpha e^{\alpha^2 t} erfc(\alpha\sqrt{t})$, and for $z \gg 1$, we have the asymptotic expansion $erfc(z) \cong \dfrac{e^{-z^2}}{z\sqrt{\pi}}\left[1 - \dfrac{1}{2z^2} + ..\right]$.

Self-consistent theory of ligand binding been assumed here. In any realistic situation, however, the number of ligands is finite, and therefore, the absorption by receptors will result in a reduction in their number and hence, the mean concentration. Nevertheless, our main conclusion that there is no depletion region for the ligand concentration close to the cell, should still be valid.

In Fig.1, we have depicted the time-variation of $\rho_s(t)$ and $\rho_r(t)$ for typical parameter values, by assuming the forms in Eq.9c for $\rho_b(t)$ and Eq.27 for $p(t)$. The total surface ligand concentration $\rho(t)$ is also shown for the same parameter values, which clearly displays the non-monotonic variation in time as argued above.

*3.4. Time-dependent association rate*
The time-dependence of the surface ligand concentration may also be framed in terms of a time-dependent association rate, which may be defined as

$$k_+^{eff}(t) = k_+ \frac{\rho(t)}{\rho_0} \tag{29}$$

The advantage of this approach, especially for numerical simulations, is that all the effects of concentration fluctuations due to ligand release and rebinding are taken care of in this new effective association rate, and no separate treatment of rebinding (eg. by using a reduced dissociation rate as in Eq.13) is needed. For the dissociation rate, we may therefore simply use the intrinsic value.

In order to study the time-dependence of the effective association rate over time-regimes of experimental interest, let us concentrate on the intermediate and late-time regimes, as defined earlier. In the intermediate regime, we have $\rho_b(t) \approx (1-\gamma)\rho_0$ and $\rho_r(t)$ is given by Eq.11. Upon combining the two, and making use of Eq.17, we find that

$$k_+^{eff}(t) \approx k_+\left[1-\gamma+\gamma\frac{(1-p_s)}{p_s}p(t)\right] \quad \text{when } p(t) \ll 1 \tag{30}$$

The late-time form is a little different, and involves the binding-corrected factor $\gamma_s$ (Eq.19).

$$k_+^{eff}(t) \approx k_+\left[1-\gamma_s+\gamma_s\frac{p(t)}{p_s}\right] \quad \text{when } p(t) \sim p_s \tag{31}$$

A compact expression that smoothly interpolates between Eq.30 and Eq.31 in the appropriate time-regimes may be obtained by defining the factor

$$\gamma(t) \equiv \frac{Nk_+(1-p(t))}{4\pi aD + Nk_+(1-p(t))}, \tag{32}$$

using which we find an approximate expression for the effective on-rate for all times.

Self-consistent theory of ligand binding

$$k_+^{eff}(t) \approx k_+ \left[1 - \gamma(t) + \gamma(t)\left[\frac{(1-p_s)}{(1-p(t))}\right]\frac{p(t)}{p_s}\right]. \tag{33}$$

Eq.33 provides a compact, although not completely rigorous (since the introduction of the factor 1-p(t) in Eq.32 is of an ad-hoc nature, and this additional time-dependence has not been included in the Laplace transform method used earlier), expression which smoothly interpolates between the correct limiting forms given by Eq.30 and Eq.31. This form may be suitable for use in computational studies of ligand-receptor interactions, where rebinding and consequent fluctuations in ligand concentration close to the surface are deemed important and may be captured at least approximately. A specific example is discussed in Section 5.

**4. Experimental implications**

In experimental studies of ligand-receptor binding, the quantity that is typically measured is the time evolution of $p(t)$, for which the effects of ligand release and rebinding on the binding kinetics are more subtle and less direct (as opposed to the ligand concentration itself). Our analysis predicts that as long as the bound fraction is sufficiently small, the binding curve may be very closely approximated to an exponential, with both the on and off-rates scaled by the same factor $1-\gamma$ (Eq.13 and Eq.14). However, closer to the steady state, the time scale of the exponential undergoes a change and switches to $T_s$, in which only the off-rate is scaled (by $1-\gamma_s$) (Eq.26). Although, there is a significant change in the surface ligand concentration, the qualitative nature of the binding curve is likely to be only subtly altered since both time scales are involved in determining the kinetics of evolution. In particular, the cross-over from Eq.18 to Eq.27 is likely to be difficult to observe in practice. This is particularly true if the equilibrium bound fraction is small, i.e., $p_s \ll 1$, for which the transient regime might well extend almost up to equilibrium. For example, in the experiments of Erickson et. al. that verified Berg-Purcell scaling [6] by effectively tuning the number of receptors, this fraction was below 0.5 in most cases. The authors also state that corrections due to the reduction in free receptor number due to binding were never observed, which lends further support to our assertion.

In order to see the differences in time evolution as predicted by Eq.13 and Eq.27 in two different time regimes, it is helpful to define the dimensionless parameter $\alpha = Nk_+/4\pi aD$. It therefore follows easily that $\gamma = \alpha/(1+\alpha)$ and $\gamma_s = \alpha(1-p_s)/[1+\alpha(1-p_s)]$. Let us also define dimensionless time as $T = k_- t$, and in terms of these new variables, Eq.13 becomes

$$p(T) = p_s[1 - e^{-\frac{T}{(1+\alpha)(1-p_s)}}] \text{ for } a^2/D \ll t, p(T) \ll 1, \tag{34}$$

and Eq.27 becomes

Self-consistent theory of ligand binding

$$p(T) = p_s[1 - e^{-\frac{T(1+\alpha p_s(1-p_s))}{[1+\alpha(1-p_s)](1-p_s)}}] \text{ for } a^2/D \ll t, p(t) \sim p_s. \tag{35}$$

Upon comparing Eq.34 with Eq.35, we find that the curves approach each other in the two limiting cases $\alpha \ll 1$ and/or $p_s \ll 1$. In these cases, the distinction between the intermediate and asymptotic regimes of evolution is unlikely to be sharp.

A precise estimation of the regime where the intermediate regime of evolution (Eq.34) crosses over to the asymptotic regime (Eq.35) may be obtained by demanding that the quadratic term in p(t) in Eq.12 is small compared to the linear term. The condition for applicability of Eq.13a (or Eq.34) then turns out to be

$$p(t) \ll \frac{1-\gamma}{\gamma(1-p_s)} \tag{36}$$

For small values of $p_s$, the difference between the asymptotic and intermediate modes of evolution is minor, as we have already remarked. Larger values of $p_s$ increase the time range of applicability of Eq.13a. In both cases, therefore, the over-all time evolution of p(t) will be dominated by the intermediate regime given by Eq.13a and 34, except for cases where $\gamma \approx 1 (\alpha \gg 1)$. In the experiments of Erickson et al. [6], the authors indicate that the best fit to the Berg-Purcell prediction was obtained with parameter values $4\pi a D \approx 4.96 \times 10^{-8}$ cm$^3$ s$^{-1}$ and $k_+ \approx 1.8 \times 10^{-13}$ cm$^3$ s$^{-1}$ for $N$ in the range $1.5 \times 10^5$ - $1.15 \times 10^6$. Upon substitution, these parameter values place $\alpha$ approximately in the range 0.5 - 4.0, and $\gamma$ in the range 0.33-0.8. The equilibrium receptor occupancy $p_s$ was reported as less than 0.5 in most of the experiments. For concreteness, let us use $p_s = 0.5$: we then find from Eq.31 that, even for $\alpha = 4.0$, the applicability of the intermediate regime is valid over time intervals such that $p(t) \ll 0.5$, and thus extends over almost the entire time regime up to saturation.

To summarize, it is not surprising that the experiments of Erickson et. al. did not notice any difference between the effective on-rate constants measured from (i) the slope of initial rise of the data curve and (ii) by fitting the entire data curve to a theoretical function. In order to see the effect predicted in the present paper, it would be necessary to do an exponential curve-fitting at both ends of the binding curve, i.e., near $t = 0$ and near $p \approx p_s$.

## 5. Application in a simple computational model

To investigate how the altered association rate expression might impact ligand binding, we simulated a simple single receptor-single ligand binding system, based on epidermal growth factor (EGF) binding parameters for concreteness. The experimental parameter values [17] can be found in Table 1, and the complete set of equations used are given in Appendix D. The primary question was whether application of Eq. 33 would lead to differences in EGF binding curves when compared to the Berg-Purcell-Shoup-Szabo effective rate constants (Eq.14) or simply using the intrinsic rate constants  Note that the dissociation rate constant was maintained at the intrinsic binding rate when coupled with Eq. 33 since effects of rebinding were pooled into that expression.

A slight difference in the fraction of receptor-ligand complexes formed as a function of time was evident when using Eq. 33, compared to the intrinsic association rate,  and negligible difference when compared to using the Berg and Purcell expressions (data not shown).  The maximal value of $\alpha$ for this system was ~0.2 , which suggests that the impact would be small here.  The differences resulting from the time-dependent association rate expression are better illustrated when the diffusion coefficient is lowered by a factor of ten and the $k_+$ is increased by a factor of five, which resulted in $\alpha \approx 10$ (Fig 2A, Table 1)  At early times, the new expression (Eq 33) leads to similar kinetics to those of Berg and Purcell, despite differences in the dissociation rate terms, but quickly leads to kinetics mirroring that of the intrinsic rates.  However, in all cases the same steady-state level of complexes was obtained.  Note that although the intrinsic rate of dissociation was used in the simulations for the dissociation rate with Eq.33,  the "effective" dissociation rate found from experiments (i.e. rate found if data was analyzed with a simple ligand-receptor binding model) would be altered from the intrinsic value due to rebinding which, in our analysis, is dealt with through the time-dependent association rate given by Eq.33

We then asked how the rate expressions would impact a more complex system where there were two ligands competing for the single receptor.  Using a second "mutant" EGF ligand where the intrinsic association rate was set equal to EGF but the off-rate differed by an order of magnitude, we simulated a competition binding study at 4ºC with equal concentration of EGF and the higher affinity mutant (1 ng/ml) (Fig 2b).  The effects are present using baseline EGF values (Table 1, data not shown) but, as with the single ligand system, are more prominent when a lower diffusion coefficient and higher association rate are used.   As expected, the level of mutant EGF-EGFR complexes exceeds those of EGF-EGFR with all three rate expression types (intrinsic values, Berg and Purcell expressions, and Eq.33).  The new expression matches the Berg and Purcell expression for early times for both mutant EGF complexes and EGF complexes but begins to track with the intrinsic rate in less than 10 min.  All three expressions resulted in similar late time (> 60 min) complex levels in agreement with the single ligand system.  There was some "overshoot" of the steady-state EGF-EGFR levels evident with all three expressions although none were large.

We next simulated conditions where there was an excess of mutant EGF (5 ng/ml mutant to 1 ng/ml EGF) and asked how the unequal concentration impacts complex formation.  Trends seen with the EGF mutant with regard to complex formation at equal concentrations added were found with the only difference being the significantly higher binding levels.  Complex formation with EGF however was somewhat different in that the "overshoot" was much more pronounced (Fig 2c).  In all cases the levels of complexes "overshot" the steady-state level with the new expression resulting in a larger "overshoot" compared to either the intrinsic rate or the Berg and Purcell expression.  This behavior was evident whether the mutant was of higher or lower affinity (data not shown). Clearly, the differences in the kinetic behavior between Eq.33 and the standard Berg-Purcell rates can be appreciable, and could be important in situations where kinetic measurements play a crucial role.

**6. Discussion**

Ligand binding to receptors on a spherical cell surface is a classic problem, and has been studied by several authors, starting with the monumental work of Berg and Purcell [2]. Although differing in details, much previous work has essentially dealt with calculating effective rate constants for the binding and dissociation of ligands [2-5]. The underlying idea was that the effects of diffusion limited transport of ligands and the non-homogeneous distribution of receptors in space (i.e., their confinement to the cell surface) could be absorbed into these effective rates. The question of the detailed nature of the kinetic binding curve was not explicitly addressed, although it seems to have been generally assumed that once the intrinsic rates were replaced by these effective rates, the standard mean-field (space-averaged) rate equation would be sufficient to describe the kinetics. This assumption was questioned by Goldstein and Dembo [10] who pointed out that this assumption is of an ad-hoc nature. However, the technique they used, although more systematic than the ad-hoc approach, is perturbative in nature and fails to predict the correct behavior in all the time regimes.

In this paper, we have developed a self-consistent stochastic approach to the binding problem, which has been built upon a similar formalism presented by us earlier to study ligand dissociation and rebinding in the presence of receptor clusters [11]. The principal idea behind this approach is to look at the problem in terms of individual ligand trajectories and derive an expression for the Green's function of ligand diffusion in the presence of the semi-absorbing cell surface. The final expressions for the cell surface ligand concentration (Eq.7 and 8) involve only macroscopic, experimentally measurable parameters like the ligand association rate and the receptor surface density.

Our results confirm the validity of the commonly used quasi-steady state approximation and the associated Berg-Purcell scaling relations over a certain well-defined time regime. However, we also predict that deviations from this relationship will be evident at larger times, close to equilibrium. In particular, we predict that the effective association rate is asymptotically the same as its intrinsic value, but undergoes a non-monotonic change between the Berg-Purcell scaling regime and this late time. The dissociation rate, however, remains a fraction of its true value at all times (beyond a short initial transient period) due to rebinding effects. Our principal prediction is that, if ligands are lost from the surface only through dissociation (i.e., provided other mechanisms like receptor-mediated internalization can be neglected), the ligand concentration at the cell surface (in fact, anywhere in the solution) at equilibrium is the same as the ligand density at infinity. However, significant variations with time leading to this point do occur, and have been fully characterized in the theory (Eq.18).

Self-consistent theory of ligand binding

Our results are of possible implications in many areas. The quasi-steady state scaling assumption for association and dissociation rates are widely used in computational modeling of biochemical reactions describing ligand-receptor interactions, and it is important to determine if the non-monotonic change in the on-rate and its eventual saturation at the intrinsic rate as predicted in this paper has a significant effect on these types of results. It would also be of interest to extend the ideas in this paper to cell-surface (two-dimensional) ligand-receptor reactions (where the ligand interacts with the receptor after absorption on the cell surface and reaching it by surface diffusion) and is an area we are currently pursuing. This could have direct relevance in the context of receptor clustering, as in, for example, lipid rafts [18]. The results might also be applicable to the question of gradient detection in chemosensory cells. Investigations in this direction are currently in progress and will be reported in the near future. In general, the self-consistent formalism that we have presented might be found to be of use in many other problems where external ligands/molecules/monomers are absorbed and released simultaneously (eg. motor proteins attaching to and detaching from microtubule filaments).

## Acknowledgements

S. G would like to thank Harish-Chandra Research Institute, Allahabad for hospitality during a summer project studentship under the Visiting Students Program. M. G would like to acknowledge useful discussions with G. I. Menon, D. Dhar, D. Chowdhury and I. Bose . This work was funded in part by NIH Grant R01 HL56200 (KFW).

## Appendix A

We compute the probability density for a ligand that starts from a radial distance $r_0$ from the center of a perfectly reflecting sphere (of radius $a$) to be present close to its surface at time $t$. For this purpose, let us first consider the general Green's function $G_0(r,t;r_0,0) \equiv Q(r,t)$, so that $Q(r,t)4\pi r^2 dr$ is the probability of finding the ligand in the volume element $4\pi r^2 dr$ at time $t$. By this definition, therefore, $G_0(r_0,t) = Q(a,t)$, and the initial condition for $Q$ is

$$Q(r,t=0) = \frac{\delta(r-r_0)}{4\pi r_0^2} \tag{A1}$$

$Q(r,t)$ satisfies the diffusion equation

$$\partial_t Q = \frac{D}{r^2} \partial_r (r^2 \partial_r Q) \tag{A2}$$

along with the reflecting boundary condition $\partial_r Q|_{r=a} = 0$ on the surface of the sphere.

Self-consistent theory of ligand binding

Let us define $u(r,t) = rQ(r,t)$, and change variables to $x = r - a \geq 0$, which transform Eq.A1 into the one-dimensional diffusion equation

$$\partial_t u(x,t) = D \frac{\partial^2 u}{\partial x^2} \tag{A3}$$

with initial and boundary conditions

$$u(x,0) = \frac{\delta(x + a - r_0)}{4\pi r_0} \text{ and } \partial_r u \big|_{x=0} = a^{-1} u(0,t) \tag{A4}$$

In order to solve Eq.A3 and Eq.A4 together, let us define the Laplace transform $\tilde{u}(p,s) = \int_0^\infty dt \int_0^\infty dx e^{-px-st} u(x,t)$. From Eq.A3, after using Eq.A4 we then find that

$$\tilde{u}(p,s) = \frac{D(p + 1/a)}{Dp^2 - s} \tilde{u}(x=0,s) - \frac{e^{-p(r_0 - a)}}{4\pi r_0 (Dp^2 - s)} \tag{A5}$$

where $\tilde{u}(x,s) = \int_0^\infty dt e^{-st} u(x,t)$. After the inverse transform $p \to x$, we find that

$$\tilde{u}(x,s) = \tilde{u}(x=0,s) \left[ \cosh(x\sqrt{s/D}) + \frac{1}{a\sqrt{s/D}} \sinh(x\sqrt{s/D}) \right] - \frac{\Theta(x - r_0 + a)}{4\pi r_0 \sqrt{Ds}} \sinh\left[\sqrt{s/D}(x - r_0 + a)\right] \tag{A6}$$

where $\Theta(x) = 1$ for $x \geq 0$ and is zero otherwise. In order to find $\tilde{u}(x=0,s)$, we use the condition that $\tilde{u}(x,s)$ should vanish as $x \to \infty$. From Eq.A6, this is found to be possible only if

$$\tilde{u}(x=0,s) = a\tilde{G}_0(r_0,s) = \frac{1}{4\pi r_0 \sqrt{Ds}} \left[1 + \sqrt{D/a^2 s}\right]^{-1} e^{-(r_0 - a)\sqrt{s/D}}, \tag{A7}$$

which leads to Eq.8.

## Appendix B

In this appendix, we will solve the pure dissociation problem, where there are no ligands in the bulk solution to start with, but a certain fraction p(0) of receptors are ligand-bound at $t = 0$. We are interested in the kinetics of $p(t)$. For simplicity, we assume that p(0)<<1, so that the effective equation for $p(t)$ in this case has the form

$$\frac{dp(t)}{dt} = -k_- p(t) + k_+ \rho_r(t) \tag{B1}$$

After Laplace-transformation, we find

$$[s + k_-]\tilde{p}(s) = p(0) + k_+ \tilde{\rho}_r(s) \tag{B2}$$

Let us now use Eq.9b for $\tilde{\rho}_r(s)$, from which it follows that

$$\tilde{p}(s) = \frac{p(0)}{s + k_-[1 - \Sigma(s)]} \text{ with } \Sigma(s) = \frac{Nk_+}{4\pi aD(1 + a\sqrt{s/D}) + Nk_+}. \tag{B3}$$

Clearly, for very early times $t \ll a^2/D$ (or $a\sqrt{s/D} \gg 1$ in Eq.B3), the factor $\Sigma \ll 1$, and the decay is exponential with the intrinsic rate itself. The interesting regime of decay is when $a\sqrt{s/D} \ll 1$, but $s >> (1-\gamma)k_-$ (the corresponding time regime is $a^2/D << t << [(1-\gamma)k_-]^{-1}$), in which case we find $\Sigma \approx \gamma$, and hence

$$p(t) \cong p(0)e^{-(1-\gamma)k_- t} \quad \text{when} \quad a^2/D << t << [(1-\gamma)k_-]^{-1} \tag{B4}$$

The decay is now exponential, but the rate has been modified on account of the frequent rebinding events.

Finally, at very late times, when $s << (1-\gamma)k_-$, we switch to a third mode of decay because the $a\sqrt{s/D}$ factor dominates over $s$. In this case, Eq.B3 becomes

$$\tilde{p}(s) = \frac{p(0)}{k_-(1-\gamma)[1+a\sqrt{s/D}]}, \tag{B5}$$

and its inversion gives

$$p(t) \sim \frac{p(0)a}{2k_-(1-\gamma)\sqrt{\pi D}} t^{-3/2} \quad \text{when} \quad t >> [(1-\gamma)k_-]^{-1} \tag{B6}$$

The power-law decay is in agreement with the result of Carslaw and Jaegar (in the context of heat diffusion), derived using `radiating' boundary condition at the surface [10,19,20].

## Appendix C

Although it is not directly relevant to the binding problem, it is nevertheless interesting to look at the density profile of ligands in the bulk solution in the steady state using our formalism. For this, we need the general Green's functions $G_N(r,t;r_0,0)$, which is again related to $G_0(r,t;r_0,0)$ by Eq.7. The latter is easily calculated using Eq.A6. We omit the details and give the result:

$$G_0(r,t;r_0,0) = \frac{e^{-(r-a)\sqrt{s/D}}}{4\pi r D(1+a\sqrt{s/D})} \tag{C1}$$

Using Eq. 4a and 4b, we now find that

$$\tilde{\rho}_b(r,s) = \frac{\rho_0}{s} \frac{4\pi r D}{4\pi r D + k_+ N e^{-(r-a)\sqrt{s/D}}} \tag{C2}$$

and

$$\tilde{\rho}_r(r,s) = k_- N \tilde{p}(s) \frac{e^{-(r-a)\sqrt{s/D}}}{(1+a\sqrt{s/D})4\pi r D + k_+ N e^{-(r-a)\sqrt{s/D}}} \tag{C3}$$

In the small s-limit, it is easy to see from Eq.C2 and Eq.C3 that the two density contributions approach the steady state values

$$\rho_b^s(r) = \rho_0 \frac{4\pi r D}{4\pi r D + N k_+(1-p_s)} \quad \text{and} \quad \rho_r^s(r) = \frac{k_- N p_s}{4\pi r D + N k_+(1-p_s)} \tag{C4}$$

The factor $(1-p_s)$ signifies the reduction in the availability of free receptors due to binding, as explained in the text.

The bulk-ligand density grows away from the surface and eventually saturates to $\rho_0$ at infinite distance from the cell, while the receptor-released ligand density falls away from the cell, as we expect. It is also easily verified that the sum $\rho_b^s(r) + \rho_r^s(r) = \rho_0$ everywhere.

## Appendix D

We set up simulations to mimic ligand binding performed at a 4ºC (i.e. no synthesis and no internalization). Ligands were added at time t=0 and all receptors were assumed to be unbound at that time. For a single ligand case, we used the equations

$$\frac{dR}{dt} = -k_f RL + k_r C \tag{A1}$$

$$\frac{dC}{dt} = k_f RL - k_r C \tag{A2}$$

$$V\frac{dL}{dt} = -k_f RL + k_r C \tag{A3}$$

where R is the number of unbound receptors, L is the ligand concentration, C represents the number of ligand-bound receptors, $k_f$ is the association rate, $k_r$ is the dissociation rate, and V is the media volume per cell.

We explicitly simulated three cases: (i) $k_f = k_+$ and $k_b = k_-$, i.e., the intrinsic rates, (ii) $k_f = (1-\gamma)k_+$ and $k_b = (1-\gamma)k_-$, the Berg-Purcell-Shoup-Szabo rates and finally, (iii) $k_f = k_+^{eff}(t)$ and $k_b = k_-$, with $k_+^{eff}(t)$ given by Eq.33.

For the two-ligand system, equations A2 and A3 are unchanged but A1 is rewritten as:

$$\frac{dR}{dt} = -k_f RL + k_r C - k_f^2 RL_2 + k_r^2 C_2 \tag{A4}$$

and, we have the additional equations,

$$\frac{dC_2}{dt} = k_f^2 RL_2 - k_r^2 C_2 \tag{A5}$$

$$V\frac{dL_2}{dt} = -k_f^2 RL_2 + k_r^2 C_2 \tag{A6}$$

where $L_2$ is the competing ligand, $C_2$ is the competing ligand-receptor complexes, and $k_f^2$ and $k_r^2$ are the association and dissociation rate constants for the competing ligand. Note that in simulations the association and dissociation rate "constants" were varied between the intrinsic rate constants, the Berg and Purcell expressions, and the new expressions, as explained above.

Simulations were run in Matlab R2006b (The Mathworks, Inc.) using the stiff ordinary differential equation solver ode15s using the backwards differentiation formulas option with an absolute tolerance criteria of 1 x $10^{-16}$.

**Tables**

| | | |
|---|---|---|
| $k_+$ | 9.7 x 10$^5$ M$^{-1}$min$^{-1}$ | intrinsic association rate constant[17] |
| $k_-$ | 0.24 min$^{-1}$ | intrinsic dissociation rate constant[17] |
| N | 4.0 x 10$^5$ #/cell | receptor density [17] |
| a | 4.0 x 10$^{-6}$ m | cell radius |
| D | 1 x 10$^{-6}$ cm$^2$/s | diffusion coefficient |
| V | 1 x 10$^{-9}$ L/cell | media volume per cell |

TABLE 1. The parameter values used in the simulations described in Appendix D. Note that plots in Fig.2 reflect variations from base parameters as indicated in text.

Self-consistent theory of ligand binding

**Figures**

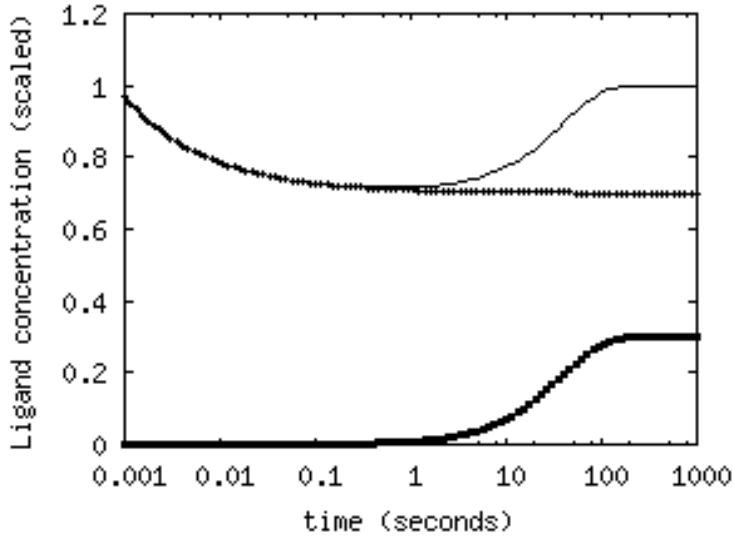

FIG 1. Total cell surface binding is dependent on both bulk and rebinding contributions. Ligand concentrations $\rho_b(t)$ (+), $\rho_r(t)$ (■) and the total cell surface concentration $\rho(t) = \rho_b + \rho_r$ (smooth line), scaled by the ligand concentration at infinite distance from the cell, versus time (plotted on a logarithmic scale) is shown. The following parameter values were used: $\gamma_s = 0.3$, $p_s = 0.5$ and $k_- = 0.01 s^{-1}$. The very short-time regime corresponding to time scales $< a^2/D \approx 10^{-3} s$ have been left out of the plot. The plateau region in the smooth line corresponds to Berg-Purcell scaling of the association rate.

Self-consistent theory of ligand binding

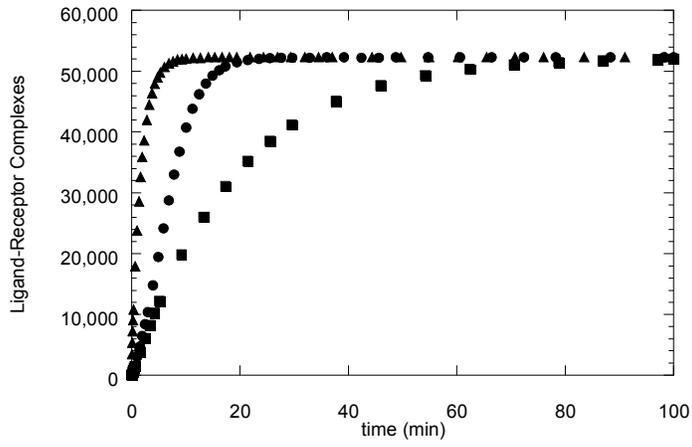

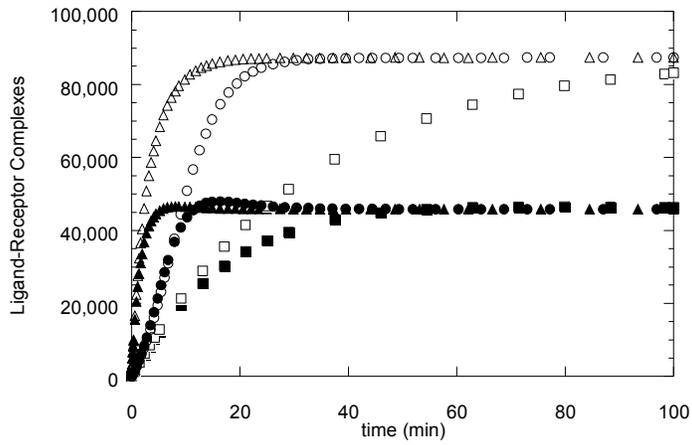

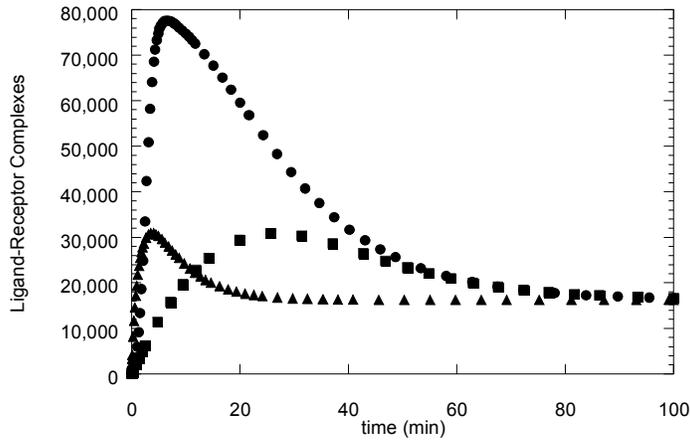

Fig 2. Altered Association Rate Impacts Ligand-Receptor Complex Formation. A. Ligand-receptor complexes for the single ligand-single receptor system are plotted as a function of time using variable association/dissociation rates: intrinsic $k_+$ and $k_-$ (triangle), Berg and Purcell analysis (square), and Eq.33 (circle), as described in Appendix D. B. Ligand-receptor complexes in the two ligand-one receptor system with equal concentrations of ligand added (EGF-EGF (filled) mutant EGF-EGFR (open)) are plotted as a function of time using variable association/dissociation ratesas described in A\. C. Ligand-receptor complexes in the two ligand-one receptor system when unequal concentrations (1 ng/ml EGF and 5 ng/ml EGF mutant) are added. Only EGF-EGFR complexes are shown and symbols are the same as listed in A. Equations and Parameters are listed in Appendix D and Table 1.